\documentclass{iau} 
\usepackage{graphicx}
\usepackage{hyperref}
\usepackage{subfig}

\title[\texttt{Mentari}: a pipeline to model the galaxy SED using semi analytic models] 
{\texttt{Mentari}: A pipeline to model the galaxy SED using semi analytic models}

\author[Dian Triani, Darren Croton, Manodeep Sinha]   
{Dian Triani,$^1$
Darren Croton$^1$
 \and Manodeep Sinha$^1$}

\affiliation{$^1$Centre for Astrophysics and Supercomputing, Swinburne University of Technology \\  PO Box 218, Hawthorn VIC 3122, Australia \\ email: {\tt dtriani@swin.edu.au}}

\pubyear{2018}
\volume{341}  
\jname{Challenges in Panchromatic Galaxy Modelling with Next Generation Facilities}
\editors{M. Boquien, E. Lusso, C. Gruppioni \& P. Tissera, eds.}
\begin{document}

\maketitle

\begin{abstract}
We build a theoretical picture of how the light from galaxies evolves across cosmic
time. In particular, we predict the evolution of the galaxy spectral energy distribution (SED) by carefully integrating the star formation and metal enrichment histories of semi-analytic model (SAM) galaxies and combining these with stellar population synthesis models which we call mentari. Our SAM combines prescriptions to model the interplay between gas accretion, star formation, feedback process, and chemical enrichment in galaxy evolution.
From this, the SED of any simulated galaxy at any point in its history can be constructed and compared with telescope data to reverse engineer the various physical processes that may have led to a particular set of observations. The synthetic SEDs of millions of simulated galaxies from mentari can cover wavelengths from the far UV to infrared, and thus can tell a near complete story of the history of galaxy evolution.
\keywords{galaxies: evolution - galaxies: stellar content - galaxies.}
\end{abstract}

\firstsection 
\section{Introduction}
The study of galaxy evolution has a long history, both theoretically and empirically. The general view of how the large-scale structure in the Universe grows includes an understanding of the virialization of dark matter and baryons, gas cooling, star formation, feedback, and mergers. Galaxy properties are driven by various possible scenarios of how such nonlinear processes occur and compete. These properties include the stellar mass, star formation history (SFH), metallicity, initial mass function (IMF), and physical state and quantity of dust and gas (\cite[Conroy 2013]{Conroy13}). \par
From the theoretical side, semi-analytic models (SAMs) have been widely used to predict the evolution of galaxies. Such models are run as a post-processing step on dark matter only N-body simulations. SAMs consist of a set of coupled differential equations, both empirically and/or physically driven, that describe each physical process believed important. Compared to a hydrodynamical simulation that solves the equations of gravity and fluid dynamics of each particle, SAMs are significantly more computationally efficient and faster at exploring the parameter space. \par
On the observation side, the spectral energy distribution (SED) of stars and interstellar medium in galaxies encode many of the fundamental aspects of galaxies that are key to studying their evolution. A large number of surveys using different methods have been conducted to measure the SEDs of galaxies, from wide field surveys of the local Universe such as the Sloan Digital Survey (SDSS, \cite[York et al. 2000]{York_etal00}), and the Two-degree Field Galaxy Redshift Survey (2dFGRS, \cite[Coless et al. 2001]{Coless01}), to the deep-field survey of the high-redshift universe such as the Deep Extragalactic Evolutionary Probe (DEEP2, \cite[Davis et al. 2003]{Davis03}). \par
To infer the properties of galaxies from the SED, an early approach involved trial-and-error techniques by combining stellar spectra until they matched the observations (\cite[Spinrad \& Taylor 1971]{SpinradTaylor71}). Following this, an automated fitting technique that incorporated physical constraints was developed (\cite[Faber 1972]{Faber72}). These methods actually matured into what has become the standard technique today, known as stellar population synthesis (SPS) modeling; it uses stellar evolution theory to constrain the range of possible stellar populations at a given age and metallicity (\cite[Tinsley 1968]{Tinsley68}). Currently, a number sophisticated models of galaxy spectral evolution have been developed for public use (e.g., \cite[Bruzual \& Charlot 2003]{BruzualCharlot03}, \cite[Conroy \& Gunn 2010]{ConroyGunn10}). \par 
This work combines the physically motivated star formation and metallicity histories from a SAM with the widely-applied method of stellar population synthesis. The predicted SED covers wavelengths from the far UV to infrared, will eventually include dust, gas and AGN physics, and thus will be able to tell a complete history of each galaxy and identify their fundamental properties. \par

\begin{figure}
\begin{center}
 \subfloat[Changing star formation efficiency]{\includegraphics[width=.5\linewidth]{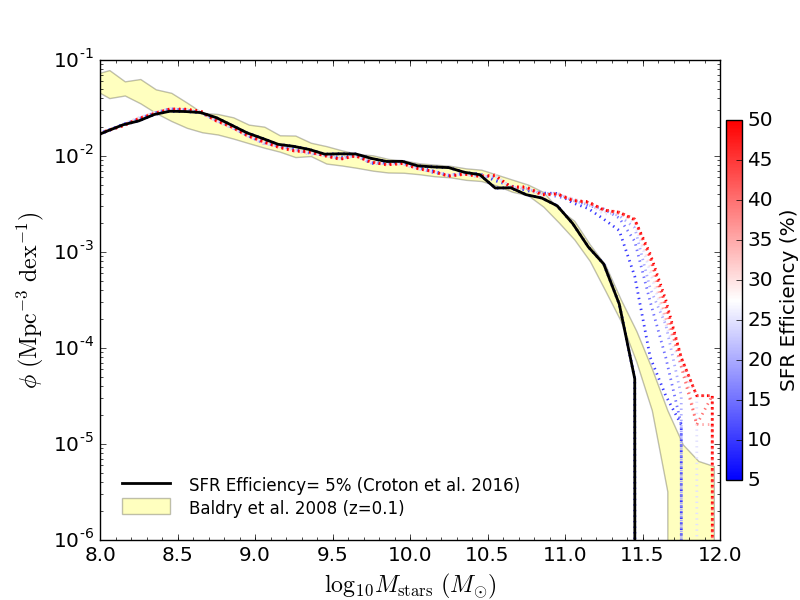}}
 \subfloat[Changing AGN feedback efficiency]{\includegraphics[width=.5\linewidth]{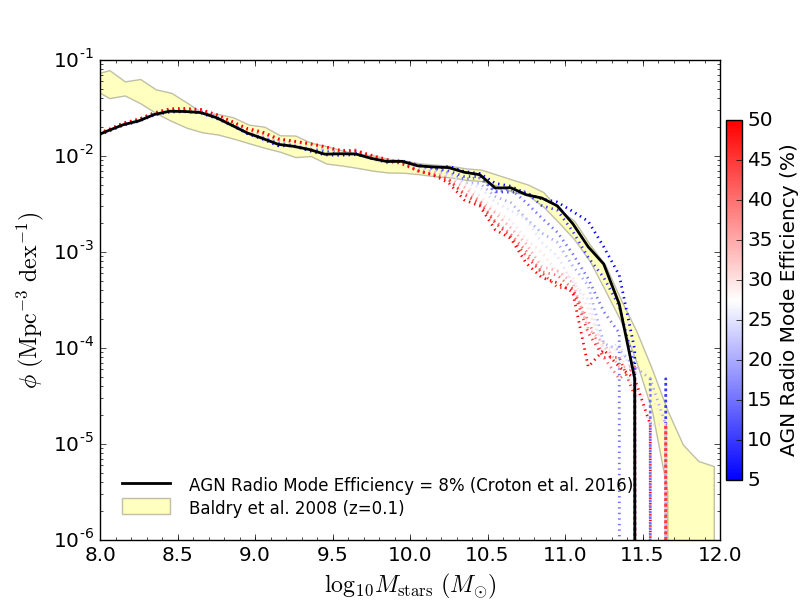}}
 \caption{Effect of changing \texttt{SAGE} parameter in the stellar mass function}
   \label{figb}
\end{center}
\end{figure}

\section{Semi-analytic model}
The galaxy model we use is called Semi-Analytic Galaxy Evolution, or \texttt{SAGE} for short (\cite[Croton et al. 2016]{Croton16}, \cite[Croton et al. 2006]{Croton06}). \texttt{SAGE} is a modular code written in C and available at \url{https://github.com/darrencroton/sage}. 
We run  \texttt{SAGE} on the Millennium simulation of \cite[Springel et al. (2005)]{Springel_etal05}. 
This cosmological simulation follows the growth and interaction of dark matter particles in a box of length 500$h^{-1}$ Mpc. There are $2160^3$ particles, each with mass $8.6 \times 10^8 h^{-1} M_{\odot}$. A dark matter halo is defined to have a minimum of 20 particles. The simulation assumes the standard WMAP1 cosmological parameters (\cite[Spergel et al. 2003]{Spergel03}). \par
\texttt{SAGE} consists of coupled equations that track the evolution of baryons inside dark matter halos, including the process of gas cooling, star formation, feedback, mergers, starbursts, and the effect of early Universe reionization. The details of the model can be found in \cite[Croton et al. (2016)]{Croton16}. When parameterizing the model, our main objective is to reproduce the stellar mass function at redshift zero. For producing stellar mass, we use the initial mass function of \cite[Chabrier (2003)]{Chabrier}. We give an example of how changing the star formation and AGN feedback efficiency can significantly change the stellar mass function in Fig.\,\ref{figb}. For this example, we run \texttt{SAGE} on the mini version of Millennium simulation, made for testing purposes. The mini version is identical to the full version but in a smaller box of side length 62.5$h^{-1}$ Mpc. We use the stellar mass function presented in \cite[Baldry, Glazebrook \& Driver (2008)]{Baldry08} as the observational benchmark in setting the model parameters. \par
\begin{figure}[t]
\begin{center}
 \subfloat[Spiral galaxy]{\includegraphics[width=.5\linewidth]{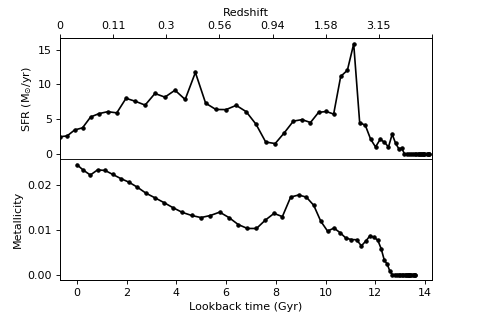}}
 \subfloat[Elliptical galaxy]{\includegraphics[width=.5\linewidth]{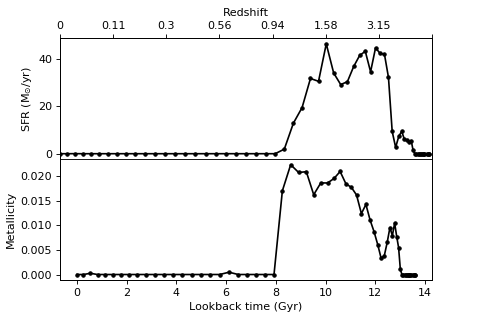}}
 \caption{Example star formation and metallicity histories of simulated SAGE galaxies (details are given in the text)}
   \label{figc}
\end{center}
\end{figure}

We compute the star formation and metallicity histories of ~36 thousand galaxies from $z = 127$ to $z = 0$ using \texttt{SAGE}. Fig.\,\ref{figc} and  shows an example of the star formation and metallicity history of a Milky Way like galaxy (morphologically spiral, stellar mass $5.86 \times 10^{10} M_{\odot}$, star formation rate $1.6 M_{\odot} yr^{-1}$) and a typical elliptical galaxy with no current star formation rate and present stellar mass of $9.3 \times 10^{10} M_{\odot}$ in the simulation. We can see the history of the spiral galaxy consists of several starbursts triggered by merger events of varying significance, and it is still undergone the star formation activity at z=0. Conversely, the elliptical galaxy had been dead since a lookback time of 8 Gyr. Using SAGE to build the star formation and metallicity histories of a large number of galaxies gives us the flexibility to customize our subject of interest. For example, it allows us to select galaxies with certain properties or history, explore the star formation activity occurring only in disk or bulge, or even look only at the intra-cluster stars around massive galaxies. The idea is to enable us to build a library of theoretical galaxies to be referenced against their observational counterparts.

\section{SED construction pipeline}
\texttt{Mentari} is an SED generation pipeline that takes semi-analytic galaxy model data as its input. The code is python-based and although still under development, has been made publicly available at \url{https://github.com/dptriani/mentari}. It has four simple functions. \par
\begin{itemize}
    \item \texttt{read\_properties(redshift, firstfile, lastfile, directory, filename)}. \par
    This function reads the properties of simulated galaxies from the \texttt{SAGE} model output.
    \item \texttt{build\_history(redshift, firstfile, lastfile, directory, filename)}. This function uses the merger tree to the correct star formation and metallicity history for each galaxy from the desired redshift. Each merger in a tree needs careful treatment because when one occurs, the old stars from the satellite galaxy will merge to the central, while the central may additionally undergo a starbursttriggered by the merger. This function will separate old stars and newly formed stars and make sure they are assigned the correct age. 
    \item \texttt{generate\_SED(lookbacktime, mass, metallicity)}. This function generates the SED from the star formation and metallicity history calculated in \texttt{build\_history}.
    \item \texttt{mab(wavelength, spectra, filter\_list, z)}. This function computes the AB magnitudes for the listed filters at the specified redshift. It will produce apparent magnitudes when the redshift is greater than zero, otherwise it assumes the user wants the absolute magnitude.
\end{itemize}
\par
To construct the SED, \texttt{mentari} reads the star formation history and metallicity evolution directly from \texttt{SAGE} and uses the \cite[Bruzual \& Charlot (2003)]{BruzualCharlot03} stellar population synthesis library, although the pipeline will be flexible and allow other libraries. This library contains spectra per unit mass of single stellar populations at ages between $1 \times 10^{5}$ and $2 \times 10^{10}$ years with metallicities that range from $0.005Z_{\odot}$ to $5Z_{\odot}$. The spectra cuurently cover panchromatic wavelengths from $91 \AA$ to $160 \mu$m. For the stellar evolution tracks and initial mass function we use Padova 1994 (\cite[Alongi et al. 1993]{Alongi93}, \cite[Bressan et al. 1993]{Bressan93}, \cite[Fagotto et al. 1994a]{Fagotto94a},
\cite[Fagotto et al. 1994b]{Fagotto94b},
\cite[Girardi et al. 1996]{Girardi96}) and \cite[Chabrier (2003)]{Chabrier03}, respectively. \par

\begin{figure}[t]
\begin{center}
 \includegraphics[width=3.2in]{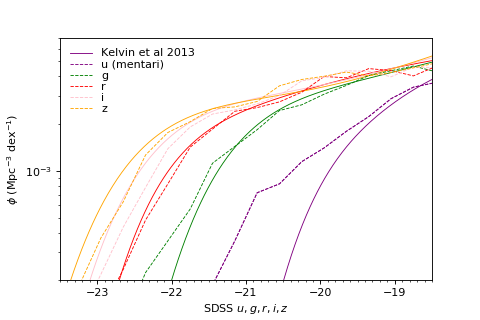} 
 \caption{Luminosity functions of simulated galaxies (thick line) and observed galaxies from \cite[Kelvin et al. 2013]{Kelvin13}. Shown are the SDSS ugriz filters, as indicated in the legend.}
   \label{figd}
\end{center}
\end{figure}

We provide a set of filters in the pipeline to convolve the SED to AB magnitude. We can create photometric datasets of galaxies at any redshift that can then be compared with observations. Using \texttt{mentari}, we construct the SED and the SDSS \textit{ugriz} photometry for ~36 thousand galaxies from our model at z=0. We compare with the luminosity function from the GAMA survey (\cite[Kelvin et al. 2013]{Kelvin13}). This is shown in Fig.\,\ref{figd}. \par
\texttt{Mentari} is currently still under development. The future enhancements will include additional stellar population models; more sophisticated dust modelling, especially for the IR part of the spectrum; quasar and radio mode AGN components to the composite SED; emission and absorption lines in the SED; and $H_I$ and $H_2$ gas components. The modular nature of \texttt{mentari} will mean that such enhancements can be included incrementally as the technical resources and scientific needs dictate.

\begin{discussion}
\discuss{Fabio Fontanot (INAF-OATs)}{Which kind of IR templates are you implementing to model the IR re-emission?}
\discuss{Answer}{No one, only pure stellar (Bruzual \& Charlot '03)}
\end{discussion}

\end{document}